\documentclass[conference]{IEEEtran}
\usepackage{amsmath,amssymb,amsfonts,xpatch}
\usepackage{cite}
\usepackage[ruled,vlined]{algorithm2e}
\usepackage{setspace}
\usepackage{tikz}
\usepackage{graphicx}
\usepackage{cool}
\usepackage{caption}
\usepackage{graphics}
\usepackage{amsmath,amssymb,amsfonts,xpatch}
\usepackage{amsthm}
\usepackage{nccmath}
 \usepackage{mathcomSTEv4}
\usepackage{acronym}
\usepackage{pgfplots}
\usepackage{subfig}
\usepackage{hyperref}
\usepackage{url}

\usepackage{tikz,pgfplots}
\pgfplotsset{width=10cm,compat=1.9}
\usetikzlibrary{calc}
\usepgfplotslibrary{fillbetween}
\pgfplotsset{compat=1.10}

\usepackage{pgfplotstable}
\usetikzlibrary{positioning}
\usetikzlibrary{pgfplots.groupplots}
\usetikzlibrary{arrows}

\acrodef{mac}[MAC]{multiple access channel} 
\acrodef{oimac}[OIMAC]{optical intensity multiple access channel} 
\acrodef{snr}[SNR]{signal-to-noise ratio}
\acrodef{awgn}[AWGN]{additive white Gaussian noise}  
\acrodef{mine}[MINE]{mutual information neural estimator}  
\acrodef{dine}[DINE]{directed information neural estimator} 
\acrodef{smile}[SMILE]{smoothed mutual information lower-bound estimator}  
\acrodef{nmie}[NMIE]{neural mutual information estimation} 
\acrodef{nit}[NIT]{neural input transformer}  

\newtheorem{theorem}{Theorem}

\newtheorem{rem}{Remark}

\newcommand{\clip}{\mathrm{clip}}

\title{  
Neural Capacity Estimators: \\How Reliable Are They?
}

\author{\IEEEauthorblockN{Farhad~Mirkarimi}
	\IEEEauthorblockA{\textit{Ryerson University}}
	\and
	\IEEEauthorblockN{Stefano~Rini}
	\IEEEauthorblockA{\textit{National Yang-Ming Chiao-Tung University (NYCU)}}
	\and
	\IEEEauthorblockN{Nariman~Farsad}
	\IEEEauthorblockA{\textit{Ryerson University}}
}
\begin{document}

\maketitle

\begin{abstract}
    Recently, several methods have been proposed for 
%
    estimating the mutual information from sample data using deep neural networks and without the knowledge of closed form distribution of the data. 
    This class of estimators is referred to as \emph{neural mutual information estimators}.
    Although very promising, such techniques have yet to be rigorously bench-marked so as to establish their efficacy, ease of implementation, and stability for capacity estimation which is joint maximization frame-work.
    %
    %
    In this paper, we compare the different techniques proposed in the literature for estimating  capacity  and provide a practitioner perspective on their effectiveness. 
    In particular, we study the performance of  \ac{mine}, 
    \ac{smile}, 
    and  \ac{dine} 
    and provide insights on 
    InfoNCE. 
    We evaluated these algorithms in terms of their ability to learn the input distributions that are capacity-approaching for the AWGN channel, the optical intensity channel, and peak power-constrained AWGN channel.
    For both scenarios, we provide insightful comments on various aspects of the training process, such as accuracy, stability, and sensitivity to initialization. 
    %
\end{abstract}


\section{Introduction}

Determining the capacity of a channel and the optimal input distribution are problem of  fundamental importance in many communication scenarios of practical relevance. 
%
The value of capacity and the optimal input distribution provide critical insight on the choice of coding rates and input constellation shape, respectively.
Generally speaking, solving the capacity problem is analytically challenging. For this reason, much research effort has been dedicated to the design of algorithms that provide approximate numerical solutions to the capacity problem. 
The scenarios in which the analytical computation of capacity is intractable and numerical approaches are necessary are cases in which 
(i) the channel transition probability expression has a complex analytical expression -- as in fiber optics
(ii) the channel transition probability is to be estimated from a set of pilot measurements.
The advent of deep neural networks (DNNs) holds the promise of mitigating the inherent complexity of the capacity estimation. 
In this paper, we try to benchmark the neural capacity estimators proposed in the literature so far and provide insights on their performance for channels of relevance in the information theory literature.
We also provide consideration on the practical issues in training the neural capacity estimators using channel samples.

\smallskip
\noindent
{\bf Relevant Literature:}

The computation of capacity using algorithmic and numerical methods most notably dates back to the well-known works of Blahut and Arimoto \cite{1054855, 1054753}. This algorithm consists of an iterative alternating maximization method using the primal and dual formulation of the capacity problem \cite[Sec. 4]{gallager1968information}.
Various authors have considered extension and refinements to this original algorithm. 
In \cite{voten}, the authors provide an algorithm to maximize mutual information in a finite state indecomposable (noise free) channel with Markov source. 
In \cite{voten1},  the approach of \cite{voten} is extended to the case of a general finite state channels.
An algorithm based on deterministic annealing is proposed for determining the capacity of a Poisson channel through gradient descent in \cite{caocapacity11}.
Finally, \cite{perm} provides a extension of Blahut-Arimoto for estimating directed mutual information and use it for estimating capacity of channels with feedback. 

Recently, a few works have considered deep learning as a tool for estimating the capacity~\cite{aharoni20_CapMemChan, aharonicomputing}, and channel coding based on mutual information maximization~\cite{wunder2019,wunder2020,tonello} and probability and geometric shaping~\cite{hoydis2020}. This new approach has been partly motivated by neural network based estimators of mutual information~\cite{oord2018representation, belghazi2018mine, 2019variational, chan2019neural, smile2020, fm0}, and can be particularly effective at estimating capacity from sample data, as deep networks can be used to  express complex relationship between channel input and outputs.  The authors of \cite{aharonicomputing} are the first to specifically consider deep learning methods for capacity computations. They focus, in particular, in determining the feedback capacity of finite state channels through reinforcement learning.
In \cite{yechannel} the approach of \cite{belghazi2018mine} is used to produce efficient joint encoder and decoders for modulation with low probability of error by producing sequences with maximum mutual information between inputs and outputs of channel. Finally, \cite{aharonicomputing} presents \ac{dine}, which relies on directed information for estimating the capacity of continuous channels with and without feedback.

\smallskip
\noindent
{\bf Contributions:}
In this work, we explore how the choice of neural network-based estimators of mutual information can affect capacity estimation. 
We focus on the problem of finding capacity of memoryless point-to-point channels with continuous alphabets. 
As in \cite{aharoni20_CapMemChan}, our approach uses neural architectures for approximating the optimal input distribution from an initial distribution by iteratively estimating the mutual information using a maximization, and then optimizing the input distribution using the estimated mutual information. 
We explore how the choice of neural network based estimators of mutual information can effect the capacity estimation. Note that while the performance of the \ac{nmie} techniques have been compared extensively for estimation of mutual information, it is not clear how they compare when estimating the capacity using this iterative double maximization approach. 

We start the paper by introducing the different neural mutual information estimators. Specifically, there are two general classes of estimators. Those that estimate the mutual information directly, such as \ac{mine} \cite{belghazi2018mine} and \ac{smile} \cite{smile2020}, and those that use entropy estimation based on reference random variables \cite{chan2019neural} such as \ac{dine} \cite{aharoni20_CapMemChan} and the lower bound based on InfoNCE \cite{oord2018representation}. Note that these entropy-based methods can be used for estimation of other information theoretic measures besides mutual information such as directed information and conditional mutual information, and that was the reason they were employed in \cite{aharoni20_CapMemChan} and \cite{fm0}.
For evaluation, we consider two continuous channels: (i) the well-known average power constraint AWGN where the optimal input is continuous, and (ii) the optical intensity channel where the optimal input has finite mass points. 
Based on our numerical evaluations we observe that for the point-to-point channels, neural mutual information estimators based on direct estimation of the mutual information give more accurate estimates of the capacity as well as the optimal input distribution. We also show that neural capacity estimation can give tighter bounds for channels with only upper and lower bounds on capacity such as the optical intensity channel.


\section{Bounds on KL Divergence}
\label{sec:Mathematical Preliminaries}

In this section, we provide a lower and an upper bound on KL-divergence: in the next section we will use these bounds to present various methods that have been proposed in the literature for estimating mutual information using neural networks.    

Let us denote the sample space of the random variable $X$ as $\mathcal{X}$, and $\mathcal{P}(\mathcal{X})$ the set of all probability measures over the Borel $\sigma$-algebra on $\mathcal{X}$.
As we consider only absolutely continuous distributions, we shall assume that  the density of the random variables (RVs) $X$ and $Y$ have a joint density $P(X,Y)$, and marginal distributions $P(X)$ and $P(Y)$.
In general, we will not explicitly indicate the independent variables of a distribution when they are clear from the context.

In the remainder of the paper, we consider three measures of information between RVs: the Kullback–Leibler (KL) divergence, mutual information (MI), and the $\chi^{2}$ divergence, defined as

%
%
\eas{
D(P || Q) &  = \int P(x) \log \f {P(x)}{Q(x)}   \diff x, 
\label{eq:KL}
\\
I(X,Y)  & = D(P(x,y) || P(x) P(y)),
\label{eq:MI}\\
\chi^{2}(P||Q) & =  \int_{}^{}\frac{(P(x)-Q(x))^{2}}{Q(x)}\diff x,
\label{eq:chi}
}
respectively, where $ 0 \log 0 / 0=0$ by definition.

We present two bounds on KL divergence, which are used in  
estimating MI. 
First bound is the Donsker-Varadhan~\cite{donsker1983} bound rewritten in the theorem below for convenience.
\begin{theorem}{\bf Donsker-Varadhan representation, \cite{donsker1983}.}
	\label{thm:KLlower}
Consider the RV $X$ and two probability measures over $\Xcal$, $P$,$Q$, then the KL-divergence can be rewritte as
	\begin{equation}\label{imj}
			D(P || Q) = \sup_{T \in \mathcal{T}} \mathbb{E}_{P}[T(X)] - \log \lb \mathbb{E}_Q\left[e^{T(X)}\right] \rb,
	\end{equation}
	where $\mathcal{T}$ is the set of all functions with finite expectations in~\eqref{imj}.
\end{theorem}

Another lower bound for KL-divergence is obtained in \cite{2019variational, fm0}.

\begin{theorem}{\bf
Tractable Unnormalized version of the Barber and Agakov (TUBA) lower
bound, \cite{2019variational, fm0}.}
\label{th:tuba}
Consider the RV $X$ and two probability measures over $\Xcal$, $P$,$Q$, then the KL-divergence is lower-bounded as
\begin{equation}
    \label{eq:tubalower}
    D(P || Q) \geq \mathbb{E}_{P}[T(X)]-\lsb \frac{\mathbb{E}_Q\left[e^{T(X)}\right]}{\alpha}+\log(\alpha)-1\rsb,
\end{equation}
where $\alpha$ is a positive constant.
\end{theorem}

%


In the next section we present different neural mutual estimation approaches which utilize the results in Th. \ref{thm:KLlower} and Th. \ref{th:tuba}.


\section{ Neural Estimation of Mutual Information (NMIE)}
\label{sec:NeuralMIestimator}

In recent years, there have been many different methods proposed for estimating mutual information using neural networks \cite{oord2018representation, belghazi2018mine, 2019variational, chan2019neural, smile2020, fm0}.
%
%
We review these methods in this section and then in the next section, we discuss how they can be employed to estimate capacity and the optimal input distribution.

These methods can be divided into groups: (i) the direct \ac{nmie} methods that evaluate the MI directly by using \eqref{eq:MI} and bounds on KL divergence, and (ii) the indirect \ac{nmie} methods that break the MI into entropy terms and estimate each of term separately using bounds on KL divergence \cite{chan2019neural, smile2020, fm0}. 
%

\subsection{Direct Mutual Information Estimation}
\label{sec:directMIestimator}

\noindent
\underline{\em MINE:}
The earliest attempt to NMIE estimation can be found in \cite{belghazi2018mine}, where the 
Mutual Information Neural Estimator (MINE) is proposed. 
The MINE is obtained from the the Donsker-Varadhan representation in Th. \ref{thm:KLlower} when a DNN is used to represent $T(X,Y)$ in \eqref{imj}. Note that here the KL divergence is between the joint distribution and the product of the marginals as in \eqref{eq:MI}, and hence the function $T(.)$ is a function of $X$ and $Y$.
It is noted in \cite{belghazi2018mine} that the optimization of $T(X,Y)$ using a DNN through stochastic gradient descent (SGD) is generally challenging. 
This is because using a naive gradient estimate over the samples of the mini-batch leads to a biased estimate of the full gradient. Exponential moving average is also proposed in \cite{belghazi2018mine}  for mitigating this problem. 
However, this leads to an MI estimator with large variance, as noted in \cite{2019variational}.
In \cite{belghazi2018mine}, the MINE is shown to be effective in preventing the MODE collapse in Generative Adversarial Networks (GANs).

\noindent
\underline{\em SMILE:}
To address the large variance of the MINE, the authors of \cite{smile2020} propose the   Smoothed Mutual Information ``Lower-bound'' Estimator (\ac{smile}). In this method, the Donsker-Varadhan representation in Theorem \ref{thm:KLlower} is rewritten as in the following theorem. 
\begin{theorem} {\bf SMILE \cite{smile2020}}
Consider the RV $X$ and two probability measures over $\Xcal$, $P$ and $Q$. The KL-divergence can then be approximated as
	\begin{flalign}
		D(P || Q) \approx \sup_{T \in \mathcal{T}} \mathbb{E}_{P}[T(X)] - \log \lb \mathbb{E}_Q\big[\clip(e^{T(X)},\tau,-\tau)\big]\rb,
    \label{eq:smile bound}
\end{flalign}
where $\clip(x,\tau,-\tau) \triangleq \max\{\min\{x,\tau\},-\tau\}$.
\end{theorem}
SMILE leverages the bound in \eqref{eq:smile bound} as the $\clip$ function is equivalent to clipping the log density ratio estimator in the interval $[-\tau,+\tau]$. Similarly to \ac{mine},  the MI can be estimated by using a neural network to represent $T(.)$, and optimizing the neural network using SGD.
The choice of $\tau$ crucially affects the bias-variance trade-off: with smaller $\tau$, the variance of the estimator is reduced at the cost of increasing its bias. \newline
%

\noindent
\underline{\em InfoNCE Contrastive Lower Bound:}
The InfoNCE bound provides a method for computing MI directly using $K$ samples of joint distribution of $X,Y$. This bound is given by
$$I(X; Y) \ge \mathbb{E}_{p^K(x,y)}\left[\frac{1}{K} \sum_{i=1}^K \log \frac{e^{\color{black}f(x_i, y_i)}}{\frac{1}{K}\sum_{j=1}^K e^{\color{black} f(x_j, y_i)}}\right].$$
where $K$ is mini-batch size and $(x_i,y_i)$ are samples of the joint distribution.


\subsection{Indirect Mutual Information Estimation}

We begin the section by presenting variational bounds on the entropy  \cite{chan2019neural}:
%
a reference (and arbitrary) distribution $Q$ over the random variable $X$ with pdf $q(x)$ is used in place of the true and unknown distribution $P$ with pdf $p(x)$. 
Using $Q$, the entropy of the random variable $X$ can be written as:
\begin{equation}\label{mes}
	h(X)=\mathbb{E}_P [-\log(q(x))]-D(P||Q).
\end{equation}
Note that the first term is the cross-entropy term $h_{\text{CE}}(P,Q)$.
Using \eqref{mes}, and i.i.d. reference random variables $X^{'}$ and $Y^{'}$, mutual information is represented as

{
	\begin{align}
		&I(X;Y) = h(X) + h(Y) - h(X,Y),  \nonumber \\
		& =D(P_{X,Y}||Q_{X^{'},Y^{'}})-D(P_X||Q_{X^{'}})-D(P_Y||Q_{Y^{'}}). \label{eq:decompose}  
	\end{align}
}%
Note that since we can choose the reference distributions to be i.i.d., the cross-entropy terms in \eqref{mes} will cancel out leaving only the KL-divergence terms.

We now present the methods that use this approach for estimating mutual information.


\noindent
\underline{\em DINE:}
The Directed Information Neural Estimator  (\ac{dine}) \cite{aharoni20_CapMemChan} estimates the directed information, which is used to compute the capacity of channels with feedback. It can also be used to estimate the mutual information and the capacity of memoryless channels.  Instead of using \eqref{eq:decompose} to decompose mutual information, \ac{dine} uses $h(Y)$ and $h(Y|X)$ and hence has two KL divergence terms to estimate instead of three in equation \eqref{eq:decompose}. Each KL divergence term is estimated using \eqref{imj} adopted from~\cite{chan2019neural}. Since the KL divergence term corresponding to $h(Y|X)$ is positive while the KL divergence term corresponding to $h(Y)$ is negative, using \eqref{imj} estimates a quantity which is not a lower bound. To support feedback and channels with memory, \ac{dine} uses a recurrent neural network to estimate the directed information and hence it is slower to train for memoryless channels without feedback. This can be easily resolved however by using fully connected layers to represent the function $T$ as is done in \cite{fm0}. %

\begin{rem}
{After extensive numerical experimentation, we have observed that InfoNCE lower bound shows a high variance (and bias) in estimation of the capacity value and optimal input distribution. 
We have also evaluated the \ac{nmie} method based on the reverse Jensen inequality in \cite{frit20SPAWC-RJE}. Our preliminary results have not shown promising results for capacity estimation using this method. Therefore, we will not include the numerical evaluation results for InfoNCE and the reverse Jensen inequality.  }
\end{rem}




In the next section we present how each of these neural mutual information estimators can be used to estimate the capacity and the optimal input distribution. Then we will compare each method numerically. 

%

\section{Neural Capacity Estimation}
\label{sec:Neural Capacity Estimation}

 \begin{figure*}[t]
     \centering
     \vspace{+0.5cm}
    \includegraphics[width=15cm]{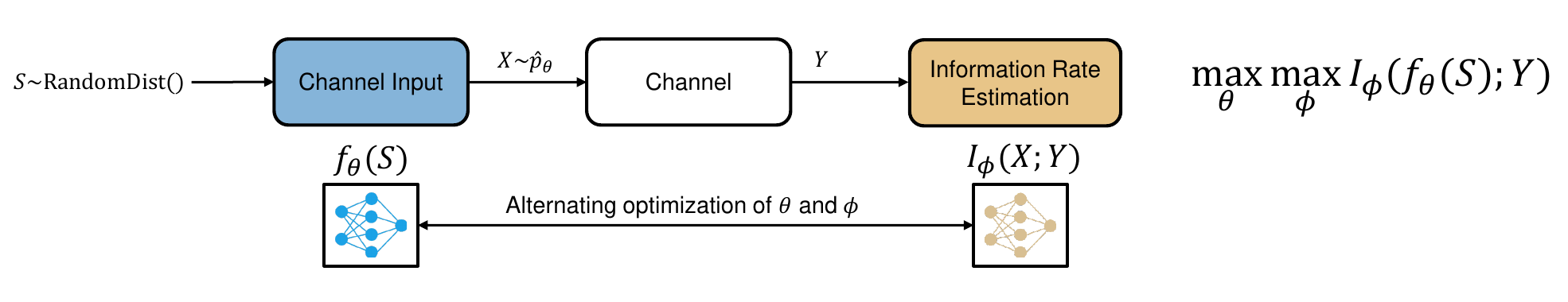}
    \caption{Joint optimization framework for neural capacity computations.}
    \vspace{-0.5cm}
    \label{fig:architecture}
\end{figure*}

In this section, we use the \ac{nmie} approach in Sec. \ref{sec:NeuralMIestimator} to estimate the capacity $\Csf$ of memoryless point-to-point channels, as described by conditional probability $P_{Y|X} \in \Xcal \times \Ycal$. The channel capacity in this case is given by 
%
\ea{
\Csf=\sup_{P \in \Pcal(\Xcal)} \ I(X,Y)
\label{eq:capacity}
}
where $X \in \Xcal$ is the channel input, $Y \in \Ycal$ is the channel output, and the supremum is taken over all continuous probability measures $P \in \Pcal(\Xcal)$ over the input alphabet $\Xcal$. The architecture we employ for the optimization in \eqref{eq:capacity} is similar to \cite{aharoni20_CapMemChan} and uses two different neural networks, the NMIE to estimate the mutual information and one called the \ac{nit} to find the capacity approaching input distribution. Specifically, the optimization in \eqref{eq:capacity} is rewritten as  
\ea{
\Csf \approx \max_\theta \max_\phi I_\phi(f_\theta(S),Y),
\label{eq:doubleMax}
}
where $I_\phi$ is an \ac{nmie} parametrized by $\phi$, and $f_\theta$ is the \ac{nit} network parametrized by $\theta$. The overall architecture is shown in Fig.~\ref{fig:architecture}.
Note that, while the performance of the \ac{nmie} techniques have been compared extensively for estimation of mutual information, it is not clear how they compare when estimating the capacity using iterative double maximization in \eqref{eq:doubleMax}.
We describe each of these network performance next. 

\smallskip

For the NMIE, any of the the four methods presented in the previous sections, as well as other NMIEs, can be used. In this work, the NMIE uses a number fully connected layers since we are considering memoryless channels. The \ac{nit} also uses fully connected layers.  The input to \ac{nit} is a standard Gaussian RV $S$ and the output of the \ac{nit} is the channel input $X$. Hence the \ac{nit} function $f_\theta$ transforms the standard Gaussian RV into any other distribution through parameters $\theta$. 

The training of \ac{nit} and \ac{nmie} is performed iteratively (similar to GANs).
Let us summarize the  training procedure as  in Algorithm \ref{alg}: a single training iteration has 3 phases. In {\bf phase 0}, the NMIE is trained exclusively for a few iterations to have the MI estimation converge to a reasonable value. 
This is performed only once at the beginning. The algorithm then enters the main optimization loop where in {\bf phase 1} the weights of the NIT network are kept constant and the NMIE network is trained and in {\bf phase 2} the weights of the NIT network are kept constant and the NMIE network is trained. The loss function used for training both networks is the negative of the estimated mutual information and is different for each of the three \ac{nmie} methods considered in this paper as described in the previous section. The main training loop continues in this fashion until the estimates $I_\theta(f_\phi(S);Y)$ converge or until a specific number of iterations are reached.

\begin{algorithm}[t]
	\caption{Neural capacity (achievable rate) estimation}\label{alg}
	\KwIn{Channel model or its GAN approximation}
	\KwOut{Estimate of the capacity or an achievable rates for the channel}
	Initialize parameters of NMIE and NIT randomly \\
	\textbf{Phase 0}: perform initial training of NMIE network using randomly generated samples \\
	\While{not converged or $\max$ iteration not reached}
	{
		Generate $B$ sample of $S$: $\{(s^{(i)})\}_{i=1}^B$ \\ 
		Generate $\{(x_i, y_i)\}_{i=1}^B$ using $x_i = f_\theta(s^{(i)})$ and channel \\
		Use NMIE to estimate:
		 $${I}_\phi(X;Y)$$  \\
		\textbf{Phase 1}: \textbf{Train NMIE} \\
		 Keep $\theta$ constant and train NMIE using stochastic gradient ascent  \\
		\textbf{Phase 2}: \textbf{Train NIT}\\
		 Keep $\phi$ constant and train NIT using stochastic gradient ascent \\
	}
	Perform final evaluation on all or subset of data \\
	\textbf{Return}: Estimated capacity
\end{algorithm}


\section{Numerical Results
}
\label{Sec:numm}


This section evaluates the proposed method on different channels with different noise models. 
We focus, in particular, on two model of broad interest:   (i) the Gaussian Additive White Noise (AWGN)  with power constraint and (ii) the optical intensity channel. 
Both channels are described by the input/output relationship
\ea{
Y &= X + N.
\label{eq:channel}
}
For (i) we have the constraint  $\Ebb[\| X\|^{2}] \leq \ep$ and $X \in \mathbb{R}$, while for (ii) we have the additional constraint  $X \in [0,A]$ for some $A \in \Rbb^+$.
In the case of the AWGN channel with an input power constraint, the capacity and the optimal input distribution are both well-known analytically. We use this channel to validate and compare the performance of each \ac{nmie} for capacity estimation. Note that while the performance of the \ac{nmie} techniques have been compared extensively for estimation of mutual information, it is not clear how they compare when estimating the capacity using the iterative method described in Algorithm~\ref{alg}.
For the case of the AWGN channel with a peak input constraint, it is known that optimal input distribution is discrete but capacity and exact optimal distribution don't have analytical formations and must be evaluated numerically. 
In the following, we shall compare the performance of the four approaches in Sec. \ref{sec:NeuralMIestimator} for first channel and a variant of second channel where $A \rightarrow \infty$. We begin the section by describing the neural network architectures used for each method and then present the performance results.

\subsection{Architectures} 
For all the experiments in this section, with the exception of \ac{dine}, \ac{nmie} network used in computation of mutual information is base on feed-forward fully connected layers. For \ac{dine}, we are using the code that was released with  \cite{aharoni20_CapMemChan}, which uses a modified LSTM. We use ReLU activation functions in all layers. Based on the method used for computation, 4 to 10 layer neural networks have been chosen for training. The hidden layers dimension is between 64 to 256 for different methods. The \ac{nit} architecture uses 5 fully connected layers with the hidden layer dimension of 64.
We have also experimented with the convolutional neural networks and no significant improvement in capacity estimation was observed.

\subsection{Data Generation and Hyperparameter Choice}
 We have used a batch size of 256 for training our algorithms with SMILE and MINE method.
All algorithms are trained in a few minutes at each SNR. The learning rate is set to $.0001$ for each of NIT and NMIE networks. We have used Adam optimizer and for SMILE method the clipping parameters $\tau$ is set to $0.2$. We use gradient clipping in all cases to avoid exploding gradients.

\subsection{Numerical Experiments}
\begin{table}[b]
\tabcolsep=0.11cm
\begin{tabular}{ |c|c|c|c|c|c|c|} 
 \hline
  SNR ($\rm dB$) & DINE & \bf{MINE}  & SMILE & True \\ 
  \hline
2 & $0.39 \pm 0.12$ & $\mathbf{0.476\pm 0.05}$ & $0.42\pm 0.09$  & 0.474  \\ 
 \hline
20  & $2.21 \pm 0.16$ & $\mathbf{2.29\pm 0.06}$ & $2.12\pm 0.07$ & 2.307  \\ 
 \hline
 40 & $4.11\pm 0.23$  &  $\mathbf{4.49\pm 0.08}$ & $4.42\pm 0.17$  & 4.605  \\ 
 \hline
\end{tabular}
\caption{Estimated capacity of AWGN channel using different methods across 10 trials.}
\label{tab:AWGN}
\end{table}

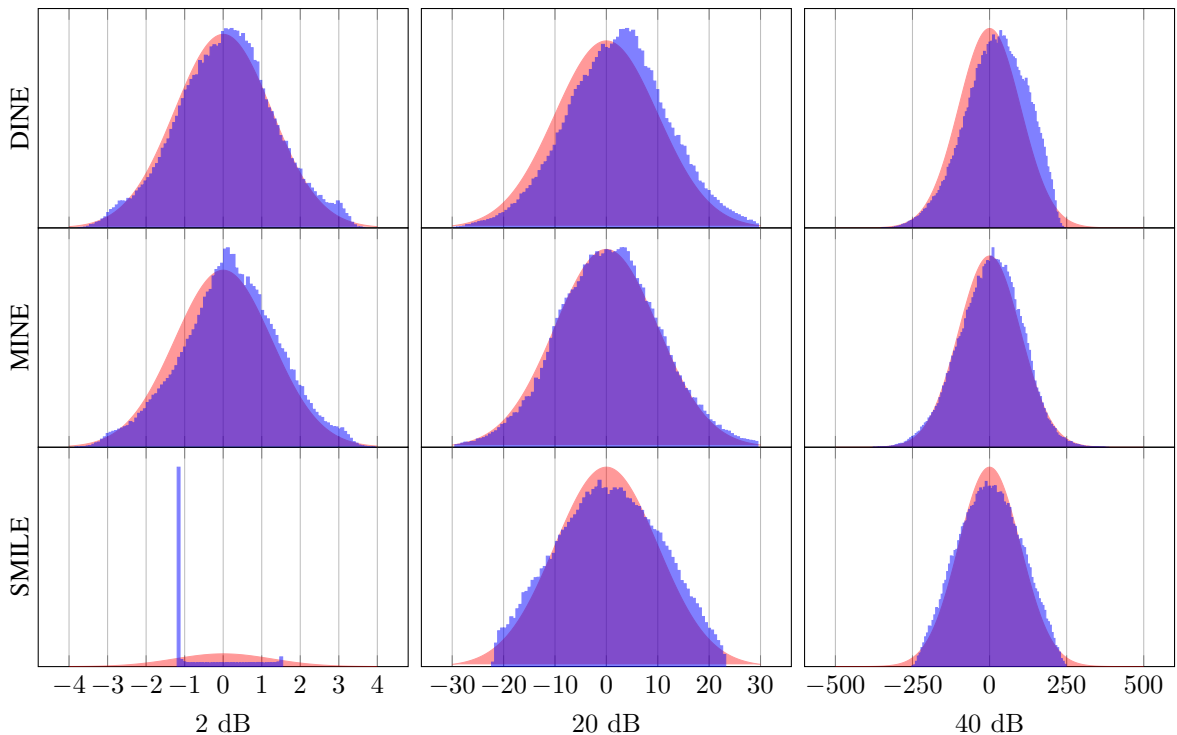
\begin{figure*}[t!]
      \centering
     \begin{tikzpicture}
    \definecolor{mycolor1}{rgb}{0.00000,0.44706,0.74118}%
    \definecolor{mycolor2}{rgb}{0.63529,0.07843,0.18431}%
    \definecolor{mycolor3}{rgb}{0.00000,0.49804,0.00000}%
    
    \begin{groupplot}[
        group style={
            group name=my plots,
            group size=3 by 3,
            xlabels at=edge bottom,
            xticklabels at=edge bottom,
            vertical sep=0pt,
            horizontal sep=5pt
        },
        height=4.5cm,
        width=6.5cm,
        ymin=0,
        ytick=\empty,
        xmajorgrids,
        ymajorgrids,
     ]
]


\nextgroupplot[ylabel=DINE,xtick={-4,-3,...,4},]
\coordinate (top) at (axis cs:1,\pgfkeysvalueof{/pgfplots/ymax});

\addplot[name path=f,red,samples=100,domain=-4:4,samples=500,fill=red!120, opacity=0.4]{(1/(1.25*sqrt(2*pi)))*exp(-(x^2/(2*1.58)))};

\addplot+[ybar interval,mark=no,fill=blue!120,draw=blue, opacity=0.5, ]
    table[col sep=comma]{./data_m/data_batch_hs-dine-awgn-2.txt};

\nextgroupplot
[xtick={-30,-20,...,30},]

\coordinate (top) at (axis cs:1,\pgfkeysvalueof{/pgfplots/ymax});
 
 \addplot[red,domain=-30:30,samples=500,fill=red!120, opacity=0.4]{(1/(10*sqrt(2*pi)))*exp(-(x^2/(2*100)))};
 
 \addplot [ybar interval,mark=no,fill=blue!120,draw=blue, opacity=0.5]
 [restrict x to domain=-30:30]
 table[col sep=comma]{./data_m/data_batch_hs-dine-awgn-20.txt };
 
\nextgroupplot[xtick={-500,-250,...,500}]

\coordinate (top) at (axis cs:1,\pgfkeysvalueof{/pgfplots/ymax});

\addplot[red,domain=-500:500,samples=500,fill=red!120, opacity=0.4]
%
{(1/(100*sqrt(2*pi)))*exp(-(x^2/(2*10000)))};
\addplot[ybar interval,mark=no,fill=blue!120,draw=blue, opacity=0.5,] 
table[col sep=comma]{./data_m/data_batch_hs-dine-awgn-40.txt};

\nextgroupplot[ylabel=MINE,xtick={-4,-3,...,4},]

    \addplot[name path=f,red,samples=100,domain=-4:4,samples=500,fill=red!120, opacity=0.4]{(1/(1.25*sqrt(2*pi)))*exp(-(x^2/(2*1.58)))};
\addplot+[ybar interval,mark=no,fill=blue!120,draw=blue, opacity=0.5]
  [restrict x to domain=-4:4]
    table[col sep=comma]{./data_m/data_batch_hs-mine-awgn-2.txt};

\nextgroupplot[xtick={-30,-20,...,30}]

\coordinate (top) at (axis cs:1,\pgfkeysvalueof{/pgfplots/ymax});
 
 \addplot[red,domain=-30:30,samples=500,fill=red!120, opacity=0.4]{(1/(10*sqrt(2*pi)))*exp(-(x^2/(2*100)))};

 \addplot [ybar interval,mark=no,fill=blue!120,draw=blue, opacity=0.5]
 [restrict x to domain=-30:30]
 table[col sep=comma]{./data_m/data_batch_hs-mine-awgn-20.txt};
 
\nextgroupplot[xtick={-500,-250,...,500}]

\addplot[red,domain=-500:500,samples=500,fill=red!120, opacity=0.4]
%
{(1/(100*sqrt(2*pi)))*exp(-(x^2/(2*10000)))};
\addplot[ybar interval,mark=no,fill=blue!120,draw=blue, opacity=0.5,] 
table[col sep=comma]{./data_m/data_batch_hs-mine-awgn-40.txt};


\nextgroupplot[ylabel=SMILE, xlabel=  $2 \ \rm dB$,xtick={-4,-3,...,4},]

\addplot[name path=f,red,samples=100,domain=-4:4,samples=500,fill=red!120, opacity=0.4]{(1/(1.25*sqrt(2*pi)))*exp(-(x^2/(2*1.58)))};;

\addplot+[ybar interval, mark=no,
            fill=blue!120,draw=blue, opacity=0.5,]
            [restrict x to domain=-4:4]
    table[col sep=comma]{./data_m/data_batch_hs-smile-awgn-2.txt};
\coordinate (bot) at (axis cs:1,\pgfkeysvalueof{/pgfplots/ymin});

\nextgroupplot[xtick={-30,-20,...,30}, xlabel=$20 \ {\rm dB}$ ]

\coordinate (top) at (axis cs:1,\pgfkeysvalueof{/pgfplots/ymax});

 \addplot[red,domain=-30:30,samples=500,fill=red!120, opacity=0.4]{(1/(10*sqrt(2*pi)))*exp(-(x^2/(2*100)))};

 \addplot [ybar interval,mark=no,fill=blue!120,draw=blue, opacity=0.5]
  [restrict x to domain=-30:30]
 table[col sep=comma]{./data_m/data_batch_hs-smile-awgn-20.txt};

\coordinate (bot) at (axis cs:1,\pgfkeysvalueof{/pgfplots/ymin});

\nextgroupplot[xlabel=$40 \ {\rm dB}$,xtick={-500,-250,...,500}]

\addplot[red,domain=-500:500,samples=500,fill=red!120, opacity=0.4]
%
{(1/(100*sqrt(2*pi)))*exp(-(x^2/(2*10000)))};
\addplot[ybar interval,mark=no,fill=blue!120,draw=blue, opacity=0.5,] 
table[col sep=comma]{./data_m/data_batch_hs-smile-awgn-40.txt};


\end{groupplot}

\end{tikzpicture}
      \caption{Comparison of different methods for computing the optimal input distribution for Gaussian channel. Each row corresponds to SNR equal to $ 2 \ {\rm dB}$, $20 \ {\rm dB}$ ,and $40 \ {\rm dB}$ respectively.}
     \label{fig:hist}
  \end{figure*}

\noindent
{\bf AWGN with average power constraint:}
%
Results for this channel are obtained using 10 separate estimation trials, where for each trial the NIT and the \ac{nmie} networks are initialized randomly. In Table \ref{tab:AWGN} we provide a comparison between the true capacity and its neural estimate for different SNR values. Specifically, we provide the average estimate across the 10 trials the standard deviation across trials.
We note that the relative accuracy decreases as the SNR grows but the loss of accuracy does not affect all methods equally.
%
%
The MINE gives a more accurate estimate of the capacity compared to other methods, while \ac{dine} seems to diverge from the true capacity at higher SNRs.  Moreover, the standard deviation of MINE is lower than all other methods suggesting that it is more robust to random initialization and has a lower variance in estimating capacity. Note that this is contrary to estimation of mutual information, where methods such as SMILE show lower variance.      
Fig.~\ref{fig:hist} shows the histogram of the \ac{nit} 
learned optimal input distribution for the DINE, MINE, and SMILE. 
%
%
We observe that at higher SNRs the \ac{mine} achieves a more accurate approximation of the optimal input distribution, i.e. Gaussian. 
%
While \ac{dine} can also learn accurate input distributions at low SNRs, at higher SNRs, it seems to deviate from the optimal input. 
%
Note that all three methods are trained for the same number of iterations here for a fair comparison. 

\noindent
{\bf Optical intensity channel:}
The capacity of optical intensity is generally unknown \cite{oimac1}, \cite{oimac2}. Since the exact capacity with closed-form solution is not available for this channel we compare our results with the bounds in \cite{oimac1}, which are some of the best known bounds in the literature.
Table.~\ref{table} shows the estimated value of the capacity and achievable rates using different \ac{nmie} methods for the optical intensity channel with only the average power constraint. 
All methods converge to values that are between lower and upper bounds. Moreover, on average the estimates of MINE have lower standard deviation compared to other methods.

Based on the numerical evaluations, we conclude that 
using \ac{nmie} methods that estimate the MI directly can result in more accurate estimates of the channel capacity and the optimal input distribution. While more investigation is needed to gain a deeper understanding of this observation, we suspect that this is because the error terms in entropy-based estimations 
compound through the learning process, resulting in less accurate estimates of MI and, consequently, capacity. Moreover, entropy-based estimators tend to be less robust to the random initialization of the networks.

\begin{table*}
\centering
\begin{tabular}{ |c|c|c|c|c|c| } 
 \hline
  SNR (db) &5& 10 & 15 & 20 \\ 
 \hline
 MINE  &$0.51\pm 0.072$ & $0.840 \pm 0.085 $ & $1.360 \pm 0.140 $& $1.810\pm 0.095$\\
 \hline
 SMILE &$0.46 \pm 0.096 $ & $0.850\pm 0.181$ & $1.373\pm 0.109$ & $1.831\pm 0.144$\\ 
 \hline
 DINE & $0.429\pm 0.097$ & $0.880 \pm 0.120$ &$1.310\pm 0.121$ &$1.730 \pm 0.110$\\
 \hline
Lower bound &0.42& 0.83 & 1.34 & 1.78\\
\hline
Upper bound  &0.99& 1.48 & 1.77 & 2.22\\
\hline
\end{tabular}
\caption{comparisons of different bounds for optical intensity channel with only average power constraint. 
}
\label{table}
\end{table*}

A repository containing the code used to derive the above results is available online here
\url{https://github.com/Farhad-Mrkm/NCE_ICC-2022}.

\section{Conclusion}\label{Sec:con}
In this paper, we investigated the ability of deep neural network (DNN) to provide an accurate estimate of the capacity of a point-to-point channel and the corresponding optimal input distribution. 
As the capacity estimation problem corresponds to a mutual information (MI) maximization problem, we consider four approaches to the MI estimation using neural networks. 
%
%
We validate the proposed approach to the capacity estimation on the AWGN channel with a power constraint and a peak amplitude constraint. 
Numerical evaluations show that all methods perform rather well at moderate SNR values, while only some methods perform well at high, or low SNRs. Specifically, we show that direct neural mutual information estimators, based on DV bound provide more accurate estimates of achievable rates and optimal input distribution for AWGN and optical intensity channel compared to other neural mutual information estimation methods.
%
%
%

\bibliographystyle{IEEEtran}
\bibliography{main}

\end{document}